\documentclass[prl,twocolumn,showpacs,aps,superscriptaddress,preprintnumbers,reprint,longbibliography]{revtex4-2}
\usepackage [dvips]{graphicx}
\usepackage{amsmath}\usepackage{amsfonts}
\usepackage{mathrsfs}
\usepackage[usenames,dvipsnames]{xcolor}
\definecolor{darkblue}{rgb}{0,0,0.6}
\usepackage{tikz}
\usepackage{ifthen}
\usepackage[colorlinks,linkcolor=darkblue,citecolor=darkblue,urlcolor=darkblue]{hyperref}

\newcommand{\br}{\text{\bf r}}
\newcommand{\be}{\text{\bf e}}

\newcommand{\bv}{\text{\bf v}}

\newcommand{\mC}{\mathcal{C}}
\newcommand{\mR}{\mathcal{R}}
\newcommand{\mL}{\mathcal{L}}
\usepackage[percent]{overpic}
\definecolor{v}{rgb}{0.33,0.67,0}
\definecolor{redUP}{RGB}{138,21,56}
\definecolor{blueUP}{RGB}{21,56,138}
\definecolor{greenUP}{RGB}{56,138,21}
\definecolor{gris}{RGB}{160,160,160}

\begin{document}

\title{Irreversible Monte Carlo algorithms for hard disk glasses:
From event-chain to collective swaps}

\author{Federico Ghimenti}

\affiliation{Laboratoire Mati\`ere et Syst\`emes Complexes (MSC), Université Paris Cité  \& CNRS (UMR 7057), 75013 Paris, France}

\author{Ludovic Berthier}

\affiliation{Laboratoire Charles Coulomb (L2C), Université de Montpellier \& CNRS (UMR 5221), 34095 Montpellier, France}

\affiliation{Gulliver, UMR CNRS 7083, ESPCI Paris, PSL Research University, 75005 Paris, France}

\author{Fr\'ed\'eric van Wijland}

\affiliation{Laboratoire Mati\`ere et Syst\`emes Complexes (MSC), Université Paris Cité  \& CNRS (UMR 7057), 75013 Paris, France}

\date{\today}

\begin{abstract}
Equilibrium sampling of the configuration space in disordered systems requires algorithms that bypass the glassy slowing down of the physical dynamics. Irreversible Monte Carlo algorithms breaking detailed balance successfully accelerate sampling in some systems. We first implement an irreversible event-chain Monte Carlo algorithm in a model of {continuously} polydisperse hard disks. The effect of collective translational moves marginally affects the dynamics and results in a modest speedup that decreases with density. We then propose an irreversible algorithm performing collective particle swaps which outperforms all known Monte Carlo algorithms. We show that these collective swaps can also be used to prepare very dense jammed packings of disks.    
\end{abstract}

\maketitle

Sampling the Boltzmann distribution in dense fluids becomes a formidable computational problem as the glass transition is approached at large density or low temperature~\cite{berthier2023modern}. If conventional methods such as molecular dynamics or local Monte Carlo algorithms are used~\cite{newman1999monte,frenkel2001understanding,krauth2006statistical}, the rapidly growing timescale characterizing the glassy dynamics also controls the sampling efficiency~\cite{barrat2023computer}. The microscopic mechanisms responsible for the dynamical slowing down continue to elude our understanding~\cite{ediger1996supercooled}. This represents a fascinating physics problem, but constitutes a major obstacle to the development of algorithms that can efficiently shortcut the slow dynamics to reach and study equilibrium states close to the glass transition. Glass-formers are a challenging benchmark for systems exhibiting a complex and rugged energy landscape, even far beyond the realm of the physical world~\cite{krzakala2007gibbs,berthier2019glassy,ciarella2023machine, kim2024densest}.

Recently, an efficient Monte Carlo algorithm was developed for size polydisperse fluids, where local Monte Carlo moves are performed in an enlarged configuration space composed of particle positions and diameters~\cite{grigera2001fast,berthier2016equilibrium,ninarello2017models}. The sequential swap of particle pairs respect detailed balance and ensures that the particle size distribution is conserved~\cite{grigera2001fast}. The resulting Swap Monte Carlo algorithm (hereafter called `Swap') allows equilibration at very low temperatures, exploiting dynamic pathways unavailable to the local dynamics~\cite{ninarello2017models}. Swap paved the way for numerous physical studies~\cite{berthier2016growing,ozawa2018random,scalliet2022thirty} and computational developments~\cite{berthier2019bypassing,parmar2020ultrastable}. Diameter dynamics can be implemented in molecular dynamics, both in thermal equilibrium~\cite{berthier2019efficient} or in gradient descent~\cite{kapteijns2019fast,brito2018theory}. For hard particles, this optimisation strategy was exploited to produce jammed packings with large stability and novel physical properties~\cite{brito2018theory,hagh2022transient,corwinref}.  

The Swap algorithm samples the Boltzmann distribution owing to reversible evolution rules obeying detailed balance. In many areas of physics and applied mathematics, it was realized that giving up detailed-balance--while preserving the target distribution--can be rewarded with sampling acceleration. Ironically, the seminal 1953 article~\cite{metropolis1953equation} by Metropolis {\it et al.} presented an algorithm to sample the Boltzmann distribution for simple fluids whose elementary moves did not, strictly speaking, satisfy detailed balance. As long as the global balance condition is satisfied by the transition rates, the target distribution is correctly sampled, even if dynamic pathways again become unphysical. 

In specific instances, it can be proved that irreversible algorithms carry out their sampling task faster than in equilibrium~\cite{chen1999lifting,diaconis2000analysis}. A successful implementation of these ideas for particle models is the event-chain Monte Carlo (ECMC) algorithm~\cite{bernard2009event} that also operates in an enlarged configuration space where irreversible collective particle translations are performed. For hard disks near their hexatic ordering transition, ECMC offers a two orders of magnitude speedup that led to a better understanding of the phase diagram~\cite{bernard2011two}. This approach was extended in various directions~\cite{michel2014generalized, michel2015event,isobe2015hard, kapfer2016cell, isobe2016applicability, michel2020forward, guyon2023necessary, bierkens2023methods}, but quantitative benchmark in dense disordered states is lacking. 

\begin{figure}[t]
    \includegraphics[width=\columnwidth]{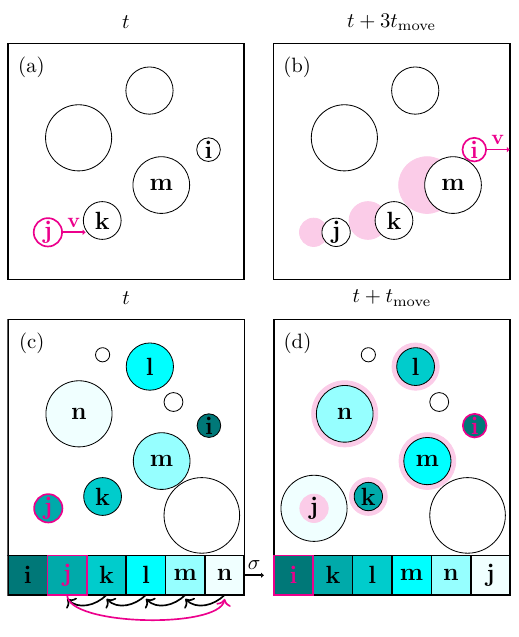}
    \caption{(a, b) Event-chain Monte Carlo algorithm: the lifted set of degrees of freedom, the active label $j$ and the speed direction $\bv$ (magenta), produce a directed translational motion of a chain of three particles. (c, d) Collective Swap algorithm: the active particle (in magenta) inflates while other particles deflate, resulting in a directed motion in diameter space (as seen at the bottom) and a collective swap of five particles.}    
    \label{fig:cswapcartoon}
\end{figure}

Here we propose, implement and benchmark irreversible Monte Carlo algorithms where collective and directed particle translations and diameter swaps are performed while maintaining global balance, see Fig.~\ref{fig:cswapcartoon}. We carefully test the respective and combined effects of these moves in {a continuously} polydisperse models of hard disks displaying glassy dynamics in equilibrium, and that can be compressed towards jamming~\cite{corwinref}. We find that the directed translational moves used in ECMC marginally affect the dynamics, with a speedup that plummets with increasing density. By contrast, irreversible collective swaps (named `cSwap') produce an opposite trend offering a comfortable gain over Swap that improves with density. Combining both types of moves in a novel algorithm (named `cSwapECMC') provides an overall computational speedup {reaching about} 40 over the conventional Swap. In addition, cSwapECMC remains extremely efficient during nonequilibrium compressions, producing jammed packings comparable to gradient descent protocols preserving the particle size distribution.

We consider a two-dimensional mixture of $N=1024$ hard disks in a periodic square box of linear size $L$ with a fixed continuous polydispersity of about $25\%$ (see SM~\cite{SM} for results on a {different polydisperse model and different system sizes}). Lengths are measured in units of the average diameter $\overline{\sigma}$, and the packing fraction is $\phi = N \pi \overline{\sigma^2}/(4L^2)$. We work in a density regime characterized by glassy dynamics, typically much beyond the one explored in \cite{bernard2009event} for monodisperse systems.

We run $NVT$ Monte Carlo simulations~\cite{frenkel2001understanding}. To compare the efficiency of various algorithms, we need to carefully define a specific unit of time, $t_\text{move}$, adapted for each case. In Metropolis Monte Carlo (MMC) dynamics, a random particle is selected uniformly, and a random displacement is uniformly drawn from a square of length $\delta$ centered around the origin. We take $\delta=0.115\overline{\sigma}$. The displacement is accepted if it creates no overlaps. One such attempt defines $t_{\rm move}$. In Swap, we randomly alternate translational moves (as in MMC) with particle swaps with probability $p_\text{swap}=0.2$. During $t_\text{move}$, two particles are randomly selected and their radii are exchanged if the swap does not create overlaps. 

In both MMC and Swap, a configuration is specified by $\mC=\{\br^N,\sigma^N\}$, encoding the $N$ particle positions and diameters. For ECMC, the phase space is lifted by two additional degrees of freedom, corresponding to the label $i$ of the active particle performing directed motion, and its direction of motion $\bv$. During a time interval $t_\text{move}$, particle $i$ travels along  direction $\bv$ until it collides with one of its neighbors, $j$. The activity label is then updated from $i$ to $j$. After a time $n t_\text{move}$, with $n$ an integer, a directed chain of $n$ particles has moved in direction $\bv$, see Fig.~\ref{fig:cswapcartoon}. To warrant ergodicity, both $\bv$ and the activity label are uniformly resampled after the total directed displacements add up to a fixed total length $\ell$ (see SM \cite{SM} for more details on the numerical implementation). Following the original choice~\cite{bernard2009event, bernard2011algorithms}, $\bv$ is uniformly resampled from $\{\be_x,\be_y\}$. Of course, ECMC can be combined with Swap, which trivially leads to a new algorithm, `SwapECMC'.

We now show how to perform directed, irreversible, collective moves in diameter space to arrive at cSwap. We define a one-dimensional array containing the particle labels in order of increasing diameters and the operators $(\mL$,$\mR)$ acting on the labels:  $\mL(i)$ returns the label of the particle immediately to the left of $i$ (with a smaller radius); $\mR(i)$ returns the label of the particle to the right (with a larger radius). During $t_{\rm move}$ we perform the following operations. A particle $i$ is uniformly selected to become active and the state of the system is described by $\mathcal{C} = \{ \br^N, \sigma^N, i\}$. We then determine the largest diameter $\sigma_j \in \sigma^N$ that particle $i$ can adopt without generating an overlap. To preserve the particle size distribution, we now perform a cascade of swaps: $\sigma_i \leftarrow \sigma_j$ (maximal authorized expansion of $i$), followed by a series of incremental deflations $\sigma_j \leftarrow \sigma_{\mathcal{L}(j)}$,  $\sigma_{\mathcal{L}(j)} \leftarrow \sigma_{\mathcal{L}^2(j)}$, $\ldots$, $\sigma_{\mathcal{L}^n (j)} \leftarrow \sigma_i$, with $n$ such that $\mathcal{L}^n (j) = \mathcal{R}(i)$, thus completing the cascade. Finally, a lifting event occurs leading to $\mathcal{C'}=\{\br^N,\sigma'^N,\mathcal{L}(i)\}$, where $\sigma'^N$ is reached after the collective swap. If $i$ is the particle with the smallest diameter, $\mathcal{L}(i)$ is the particle with the largest diameter. To warrant ergodicity, we perform with probability $1/N$ a uniform resampling of the lifting label. Finally, ECMC can be combined with cSwap, leading to a fully irreversible algorithm, `cSwapECMC'. The invention and implementation of irreversible and collective swap moves is our main algorithmic development. While cSwap is broadly applicable for any particle size distribution, its efficiency should be optimal for continuous distributions, or discrete ones with a large number of families. For bidisperse models, cSwap remains rejection-free and irreversible, but loses its collective character.

We must prove that the stationary state of the cSwap dynamics is the Boltzmann distribution, i.e. $\pi_{ss}\left(\{\br^N,\sigma^N,i\}\right) = \pi_\text{\tiny{B}}(\br^N,\sigma^N)\nu(i)$, where 
$\pi_{\text{\tiny{B}}} = [ \int d\sigma'^N \pi_{\text\tiny{B}}(\br^N,\sigma'^N) ]^{-1} \equiv Z^{-1}$ is the Boltzmann distribution for a system of polydisperse hard disks for a fixed set of positions ${\bf r}^N$, and $\nu(i)= N^{-1}$ is the uniform distribution for lifting. Denoting by $p(\mathcal{C}\to\mathcal{C}')$ the transition probability from $\mathcal{C} = \{\br^N,\sigma^N,i\}$ to $\mathcal{C}' = \{\br^N,\sigma'^N,i'\}$, we must prove that the stationarity condition
\begin{equation}\label{eq:GB}
    \sum_{\mathcal{C'}}\pi_{ss}(\mC')p(\mC' \to \mC) = \pi_{ss}(\mC)
\end{equation}
is satisfied by $\pi_{ss}=\pi_B/N$. The left-hand side is decomposed into label resampling and collective swaps: 
\begin{equation}\label{eq:p_cSwap}
    p(\mC' \to \mC) = \frac{1}{N^2}\delta_{\sigma'^N,\sigma^N} + \left(1 - \frac{1}{N} \right)\delta_{\mC',\mC^*}, 
\end{equation}
where $\mC^*$ is the configuration reaching $\mC$ after a cSwap move (we show below that $\mC^*$ exists and is unique). Substituting \eqref{eq:p_cSwap} into \eqref{eq:GB}, using the definition of $\pi_{ss}$ and $\sum_\mC = \sum_j \int d\sigma^N$, we get
\begin{equation}
    \frac{1}{N}\pi_\text{\tiny{B}}\left(\br^N,\sigma^{N}\right) + \left(1-\frac{1}{N}\right)\pi_\text{\tiny{B}}\left(\br^N,\sigma^{*N}\right) = \pi_\text{\tiny{B}}\left(\br^N,\sigma^{N}\right) . \nonumber
\end{equation}
Since for hard disks, $\pi_\text{\tiny{B}}$ is uniform over allowed configurations, stationarity is proven. Finally we construct the configuration $\mC^* = \left\{ \br^N,\sigma^{*N},i^*\right\}$ that will reach $\mC = \left\{ \br^N,\sigma^{N},i \right\}$. We first transform $\sigma_{\mR (i)} \leftarrow \sigma_{\mR^2 (i)}$ if the change does not generate any overlap. We then repeat this operation for $\mR^2 (i)$, $\mR^3 (i)$, etc. After $n$ iterations, either the transformation $\sigma_{\mR^n (i)} \leftarrow \sigma_{\mR^{n+1}(i)}$ is no longer allowed, or the largest particle is reached. 
When $n$ is reached, we set $i^*=\mR^{n}(i)$ and transform $\sigma_{\mR^{n}(i)} \leftarrow \sigma_{\mR (i)}$. The resulting configuration defines $\sigma^{*N}$, as directly verified by performing a cSwap move on $\mathcal{C}^*$.

\begin{figure}[t]
\includegraphics[width=\columnwidth]{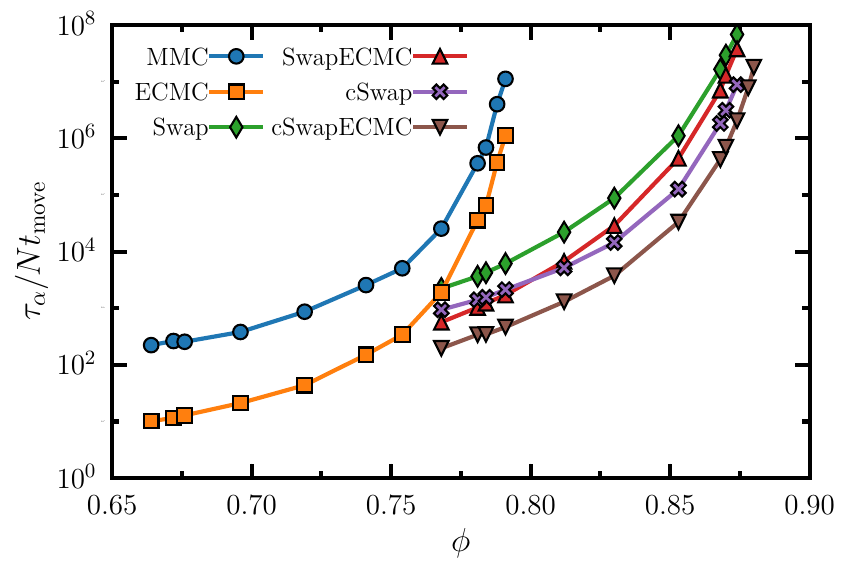}
\caption{Equilibrium relaxation time of six different Monte Carlo algorithms as a function of packing fraction. MMC and the faster ECMC fall out of equilibrium much before the four swap algorithms. The large speedup offered by Swap can be further improved using irreversible MC moves, cSwapECMC providing a further speedup of about 40 near $\phi=0.88$. \label{fig:compare_tau}}
\end{figure}

The above reasoning establishes the stationarity of the Boltzmann distribution. The general proof of ergodicity of the algorithm, as obtained for ECMC~\cite{monemvassitis2023pdmp}, is left for future work. As a test, we computed the stochastic matrix associated to the cSwap algorithm for a small system of $N=4$ hard disks (see \cite{SM}) and analytically confirmed ergodicity in that case. For larger systems, we support our claim of ergodicity by extensive numerical tests of correct sampling using cSwap, as compiled in SM~\cite{SM}.

We run simulations comparing MMC, ECMC, Swap, SwapECMC, cSwap and cSwapECMC for increasing packing fractions. After careful equilibration, we measure a representative time correlation function for $2d$ glass-formers, namely the time autocorrelation of the global hexatic order $C_\psi(t)$, and define the structural relaxation time $\tau_\alpha$ from $C_\psi(\tau_\alpha)=1/e$~\cite{SM}. For each algorithm, we collect the evolution of the correlation time $\tau_\alpha(\phi)$ measured in units of $N t_\text{move}$ in Fig.~\ref{fig:compare_tau}. The most costly part of Monte Carlo moves is the overlap detection involving a sum over neighbors. Since one such sum is needed over the time $t_{\rm move}$ in each algorithm, the comparison in Fig.~\ref{fig:compare_tau} accurately describes CPU times~\cite{SM}. 

\begin{figure}[t]
\includegraphics[width=\columnwidth]{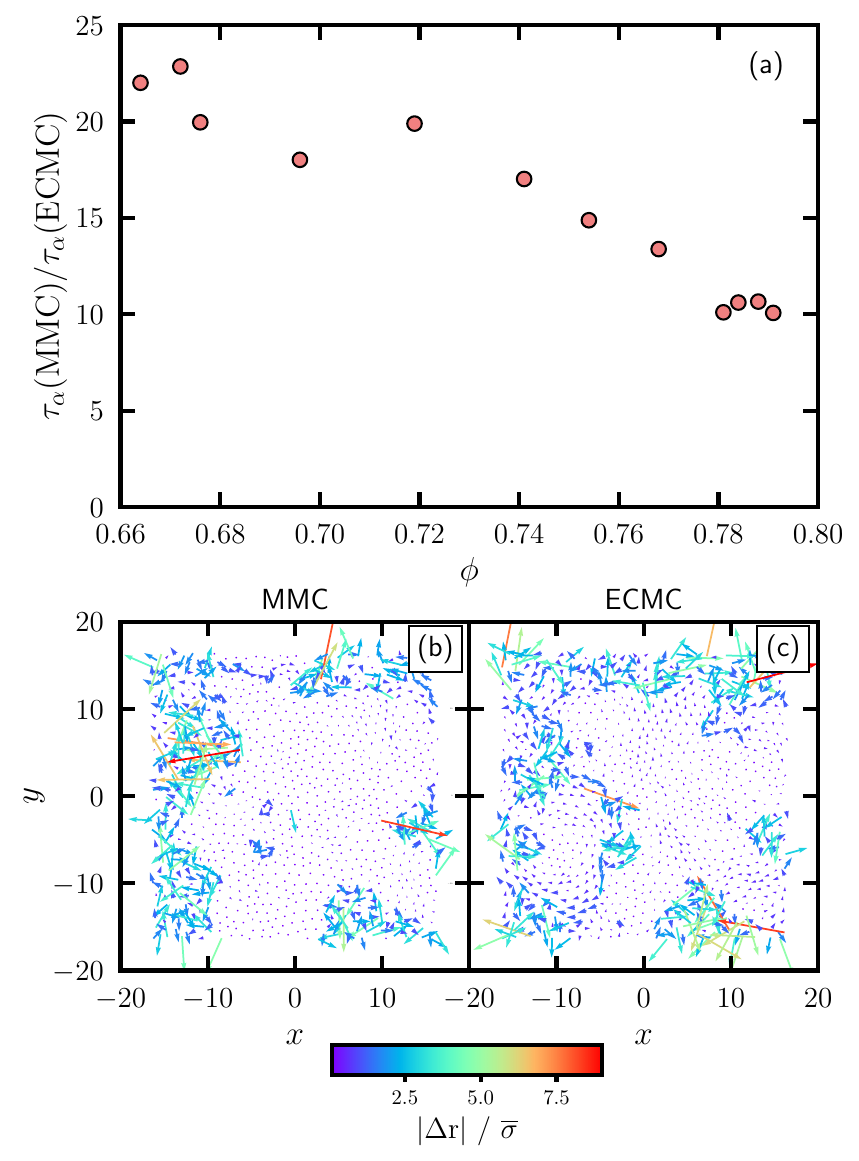}
\caption{(a) The speedup offered by ECMC over MC decreases with density. (b, c) Comparison of the displacement field relative to the center of mass after a time comparable to the relaxation time starting from the same initial condition at $\phi=0.79$ using MMC ($t=4.6 \times 10^6 Nt_\text{move}$) or ECMC ($t=2.2 \times 10^5 N t_\text{move}$). Despite different dynamic rules, both algorithms follow similar dynamic pathways.} 
\label{fig:displacement}
\end{figure}

Each algorithm displays hallmarks of glassy dynamics, and we follow for about 5 decades the slowing down. The algorithms are split into two families, depending on the presence of swap moves. MMC and ECMC only contain translations and equilibration becomes difficult above $\phi \approx 0.79$. Yet, ECMC clearly outperforms MMC throughout the entire density range, but the edge of ECMC over MMC is lost as $\phi$ increases. This is demonstrated in Fig.~\ref{fig:displacement}(a), which shows that the ratio of their relaxation times decreases from $\approx 22$ in the fluid, down to $\approx 10$ near $\phi=0.79$. This suggests that the irreversibility introduced by the directed chain moves does not help the system to discover new, faster pathways across configuration space. This interpretation is confirmed by the snapshots in Figs.~\ref{fig:displacement}(b, c) showing particle displacements with respect to the system's center of mass from the same initial condition, using either MMC or ECMC. Despite the very different particle moves in both dynamics, the long time relaxation proceeds along a similar path. A similar conclusion was recently reached for systems submitted to transverse forces~\cite{ghimenti2023sampling}. 

\begin{figure}[t]
\includegraphics[width=\columnwidth]{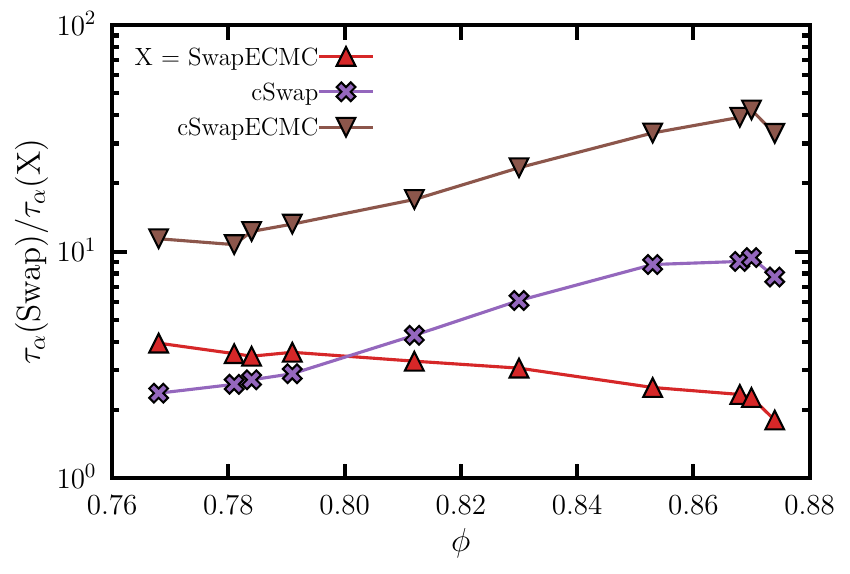}
\caption{Acceleration provided by three novel algorithms (SwapECMC, cSwap, cSwapECMC) with respect to conventional swap Monte Carlo. In cSwapECMC, the combination of collective swaps and chain moves provides the fastest algorithm with a speed up increasing with $\phi$ and reaching 40. \label{fig:cSwapwins}}
\end{figure}

By constrast, the four algorithms employing particle swaps sample the Boltzmann distribution much faster than MMC and ECMC and only become inefficient near $\phi \approx 0.88$, see Fig.~\ref{fig:compare_tau}. All algorithms thus display a dramatic speed up compared to MMC and ECMC. Using Swap as a reference, we again observe that the introduction of translational chains in SwapECMC provides a modest acceleration over conventional Swap of about 5 at $\phi=0.77$, decreasing to about 2 at the largest density (see Fig.~\ref{fig:cSwapwins}). Therefore, coupling Swap to ECMC is not very  helpful. The situation is more favorable when collective swap moves are introduced, as the speedup offered by the irreversibility in cSwap now grows with density, as demonstrated in Fig.~\ref{fig:cSwapwins}, to reach a factor about 10 near $\phi=0.88$ over Swap. These results suggest that it is useful to combine cSwap and ECMC into cSwapECMC, where both translational and diameter moves are now collective and irreversible. Getting the best of both types of moves, cSwapECMC now offers a comfortable speed up over Swap that increases from 10 to about 40 at the largest packing fraction studied, clearly outperforming the swap Monte Carlo algorithm.

\begin{figure}
\includegraphics[width=\columnwidth]{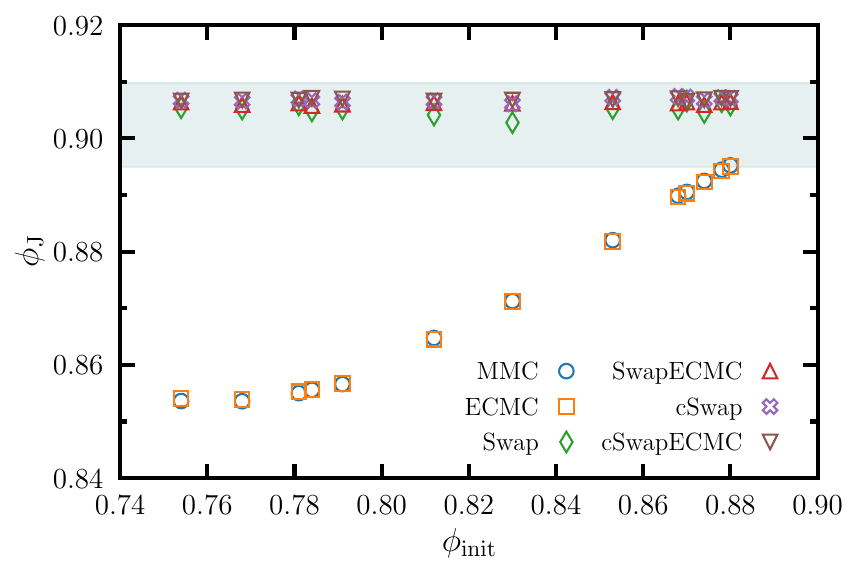}
\caption{Jamming packing fractions $\phi_J$ obtained after nonequilibrium compressions using different Monte Carlo algorithms, starting from fluid configurations equilibrated at $\phi_\text{init}$. The four swap algorithms reach very large $\phi_J$, nearly independently of $\phi_{\rm init}$. The light blue band covers the range of densities obtained using augmented gradient-descent techniques~\cite{corwinref}.}
\label{fig:phiJ_vs_phiinit}
\end{figure}

An interesting avenue for our algorithms is the production of jammed disk packings, which are typically produced using specific nonequilibrium compression protocols~\cite{lubachevsky1990geometric,ohern2002random}. Using conventional MMC for compressions, the jamming packing fraction $\phi_J$ can be reached using $NPT$ Monte Carlo. The simplest protocol starts from an equilibrium hard disk configuration at $\phi_{\rm init}$, before suddenly turning the pressure to infinity~\cite{berthier2009glass}. At long times, the packing fraction saturates to a value $\phi_J$, which is an increasing function of $\phi_{\rm init}$~\cite{chaudhuri2010jamming}. This is confirmed in Fig.~\ref{fig:phiJ_vs_phiinit}, where the range $\phi_J \sim 0.855-0.895$ is covered. Very similar results are obtained using ECMC during compressions, see Fig.~\ref{fig:phiJ_vs_phiinit}. Note that the preparation of equilibrium configurations for $\phi_{\rm init} > 0.79$ requires particles swaps~\cite{ozawa2017exploring}, which are no longer used during compressions. Interestingly, introducing swaps during compressions from the same range of initial conditions leads to jamming densities that are considerably larger, $\phi_J \approx 0.904-0.906$ (Fig.~\ref{fig:phiJ_vs_phiinit}). At the time of writing, such large packing fractions have only been obtained using gradient descent algorithms simultaneously optimizing diameters and positions to more efficiently pack the particles, followed by geometric triangulation methods~\cite{corwinref}. That similar performances can be reached using cSwap suggests that these nonequilibrium algorithms in fact explore pathways similar to the ones allowed by swap moves. In addition, the very weak dependence of $\phi_J$ on $\phi_{\rm init}$ rationalizes the surprising efficiency of augmented gradient-descent algorithms. A {major advantage} of cSwap is that the particle size distribution is {strictly} conserved, rather than annealed, during the compression.

Using swap and event-chain Monte Carlo as stepping stones, we demonstrate that simple Monte Carlo algorithms with increasing efficiency can be devised, that provide a set of improved computational tools to more efficiently equilibrate deep glassy states, prepare more stable configurations, with lower configurational entropy, thus approaching closer the putative Kauzmann transition. In order to become new standards, the cSwap algorithm and its derivatives proposed here need to be pushed in several directions. A first encouraging result is the successful scaling of their performances with system size, see \cite{SM}, in line with results for ECMC and Swap. A less obvious direction is the application to three dimensions, which is the subject of on-going efforts, again with encouraging preliminary results. A third direction concerns the application to glass-former with soft potentials. Swap performances do not decrease with continuous potentials~\cite{ninarello2017models}, and some extensions of ECMC to continuous potentials were successful~\cite{michel2014generalized, nishikawa2023liquid}. Future work should develop extensions of cSwap for glass-formers with continuous potentials to extend the range of applicability of irreversible Monte Carlo methods in the field of supercooled liquids. All these perspectives directly follow from our work; they should help the development of efficient, versatile and simple to implement sampling methods for disordered systems with a complex free energy landscape. 

\acknowledgements

We thank E. I. Corwin and V. Bolton-Lum for providing us the hard disks system studied in this work, and carefully explaining their work. We acknowledge the financial support of the ANR THEMA AAPG2020 grant, along with several discussions with G. Biroli, W. Krauth, J. Kurchan, M. Michel and G. Tarjus. 

\bibliography{biblio-cswap}

\end{document}


\title{Supplementary Material for ``Irreversible Monte Carlo algorithms for hard disk glasses: From event-chain to collective swaps"}

\author{Federico Ghimenti}

\affiliation{Laboratoire Mati\`ere et Syst\`emes Complexes (MSC), Université Paris Cité  \& CNRS (UMR 7057), 75013 Paris, France}

\author{Ludovic Berthier}

\affiliation{Laboratoire Charles Coulomb (L2C), Université de Montpellier \& CNRS (UMR 5221), 34095 Montpellier, France}

\affiliation{Gulliver, UMR CNRS 7083, ESPCI Paris, PSL Research University, 75005 Paris, France}

\author{Fr\'ed\'eric van Wijland}

\affiliation{Laboratoire Mati\`ere et Syst\`emes Complexes (MSC), Université Paris Cité  \& CNRS (UMR 7057), 75013 Paris, France}

\maketitle

\section{A model with a different polydispersity}\label{sec:different_model}

\begin{figure}[h]
    \includegraphics[width=0.6\textwidth]{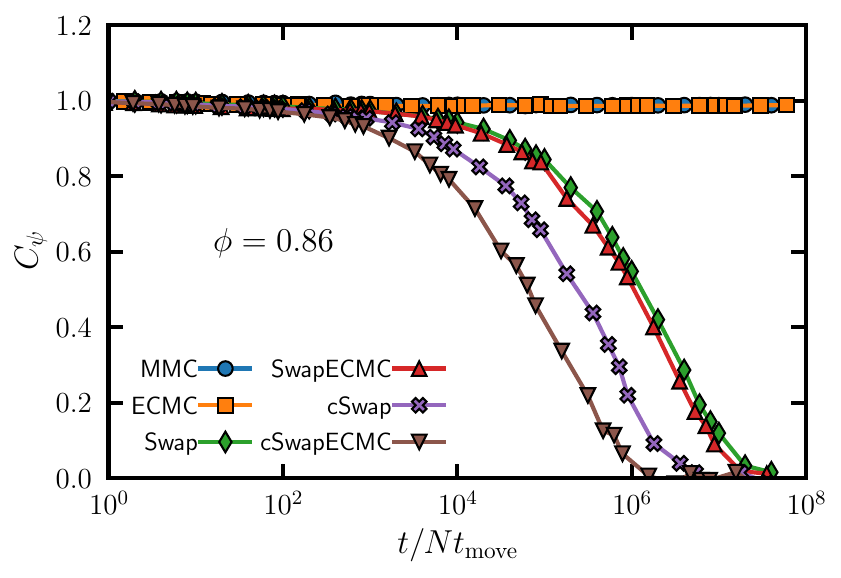}
    \caption{Time dependence of the hexatic correlation function for a system of $N=1024$ particles with a power law distribution of the diameters. Time is measured in units of $Nt_\text{move}$, and the efficiency of the algorithms that involve swap moves is comparable to the results shown in Fig.~2 of the main text near $\phi\approx 0.85$.}
    \label{fig:C6model2}
\end{figure}

To show that the efficiency of the cSwap algorithm does not depend on the specific polydispersity of the model studied in the main text, we study the relaxation dynamics of a second model. It consists of $N=1024$ hard disks with a diameter distribution following a power law, $\pi(\sigma) \propto \sigma^{-3}$, with $\sigma_\text{min}\leq\sigma \leq \sigma_\text{max}$, as studied earlier~\cite{berthier2019zero}. We recall that the units of length are chosen so that $\overline{\sigma}=1$. We have chosen the support of the diameter distribution such that the polydispersity is $\Delta=23\%$. We ran $NVT$ simulations using the six different algorithms studied in the main text at a large packing fraction, $\phi=0.86$, where we measured the time decay of $C_\psi$. The results are shown in Fig.~\ref{fig:C6model2}. The hierarchy of speedups obtained by the different algorithm, and their relative values, is comparable to the results reported in the main text for a similar value of $\phi$. 

\section{Diameter distribution of the hard disk system}

\begin{figure}[h]    
\includegraphics[width=0.5\textwidth]{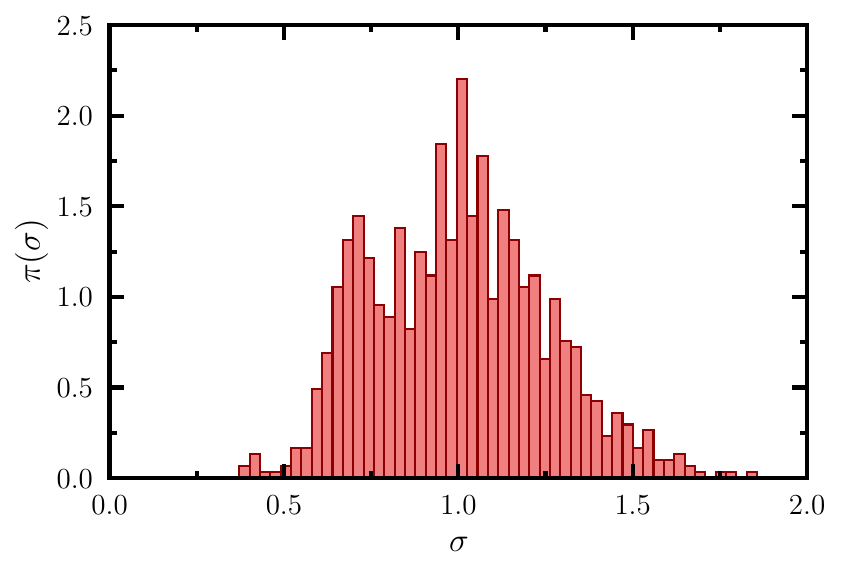}
\caption{Particle size distribution $\pi(\sigma)$ for the system investigated in this work.}
\label{fig:histodiameter}
\end{figure}

Figure~\ref{fig:histodiameter} shows the probability distribution function of the diameters $\sigma$ for the polydisperse mixture of hard disks studied in the main text. The polydipersity of the system, defined as $\Delta\equiv\frac{\sqrt{\overline{\sigma^2} - \overline{\sigma}^2}}{\overline{\sigma}} \approx25\%$. This diameter distribution results from a gradient descent protocol in an extended space composed of particles positions and radii, which will be detailed in \cite{corwinref}, which represents an improvement over earlier versions~\cite{hagh2022transient}. 

\section{Event-Chain Monte Carlo}

We recall the implementation of the Event-chain Monte Carlo (ECMC) dynamics. We resort to the so-called `straight' version, as described in \cite{bernard2009event, bernard2011algorithms}.

\begin{figure}
    \centering
    \includegraphics[width=0.6\textwidth]{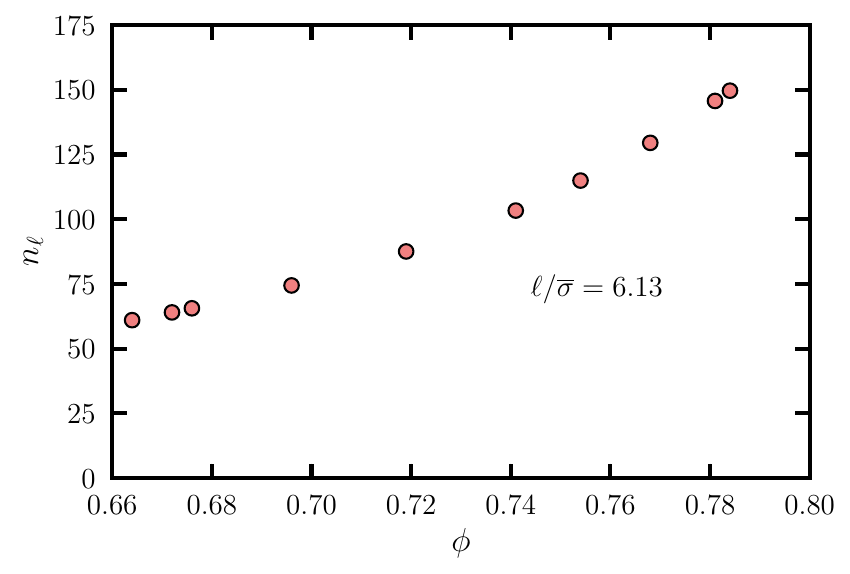}
    \caption{Average number of particles $n_\ell$ participating in a chain of fixed length $\ell$ as a function of the packing fraction $\phi$.  }
    \label{fig:supp_ecmc}
\end{figure}

The state of the system is described by a configuration $\mC=\{\br^N,\sigma^N, i, \bv\}$ where we store the position of the $N$ particles $\br^N$, their diameters $\sigma^N$, the label of the active particle $i$ and the direction of self propulsion $\bv$, a two-dimensional vector of unit norm. In the mathematical literature on Markov chains, $i$ and $\bv$ are called lifting degrees of freedom, and they govern the nonequilibrium dynamics of the system. An ECMC move consists in moving the active particle $i$ along the direction $\bv$, until an event, i.e. a collision with another particle, whose label is denoted as $j$, occurs. The distance $\delta\ell_{ij}$ traveled by $i$ during a move is thus determined by the equation
\begin{equation}
    \delta\ell_{ij} = \br_{ji}\cdot\bv - \sqrt{\sigma_{ij}^2 - \left(\br_{ji}\cdot\bv_\perp\right)^2} ,
\end{equation}
where $\br_{ji}=\br_j-\br_i$ is the vector joining particle $i$ with particle $j$, $\sigma_{ij}\equiv \frac{\sigma_i+\sigma_j}{2}$ is the effective diameter, and $\bv_\perp=(v_y,-v_x)$ is the direction orthogonal to $\bv$. In practice, in the simulations the inter-particle distance $\br_{ij}$ is computed according to the minimum image convention~\cite{frenkel2001understanding} to keep into account periodic boundary conditions, and the particle $j$ is identified through an event-driven scheme, by minimizing $\delta\ell_{ik}$ among all possible target particles $k$, i.e. $j=\arg\min_k{\delta\ell_{ik}}$. After an event occurs, a lifting move is performed: the activity label changes from $i$ to $j$. In the next ECMC move particle $j$ will perform directed motion along the direction $\bv$. The composition of several ECMC moves builds up a chain of particles performing directed motion. When the displacement performed by the active particles add up to a fixed parameter $\ell$, which fixes the length of the chain, the activity label and the self propulsion direction are resampled uniformly in their domain of definition, which are respectively the set $\{1,2,\ldots,N\}$ for the activity label and the set $\{\be_x,\be_y\}$ for the self-propulsion, for the case of straight ECMC. As the density of the system increases, more and more particles participate to the chain. This is demonstrated in Figure~\ref{fig:supp_ecmc}, which shows $n_\ell$, the average number of particles in a chain as a function of the packing fraction $\phi$ for a fixed value of $\ell$.

\section{Hexatic correlation function}

We recall the definition of the time correlations for the hexatic order parameter, which have been extensively used in the past to monitor the relaxation of two-dimensional glassy systems \cite{flenner2015fundamental}. The hexatic correlation function $C_\psi(t)$ is defined as 
\begin{equation}\label{eq:Cpsi}
    C_\psi \equiv \frac{\left\langle \psi^*(t) \psi(0)\right\rangle}{\left\langle \lvert \psi(0)\rvert^2\right\rangle} ,
\end{equation}
where $\psi(t)$ is the global hexatic order parameter. It is defined as a sum of local terms $\psi(t) \equiv \frac{1}{N}\sum_{i=1}^N \phi_{6,i}(t)$ where the local hexatic order parameter for particle $i$,  $\phi_{6,i}(t)$ is defined as 
\begin{equation}
     \phi_{6,i}(t) \equiv \frac{1}{n_i} \sum_{j=1}^{n_i} e^{6i\theta_{ij}} ,
\end{equation}
with the sum running over the $n_i$ neighbors of particle $i$, defined through a Voronoi tessellation, constructed using the freud libary \cite{RAMASUBRAMANI2020107275}, and $\theta_{ij}$ is the angle between the vector $\br_{ij} \equiv \br_i-\br_j$ and the vector $\be_x$. From the time decay of $C_\psi$ we define the relaxation time $\tau_\alpha$, such that $C_\psi(\tau_\alpha) = e^{-1}$.

\section{Ergodicity of cSwap in an $N=4$ system}

Here we prove the ergodicity of the cSwap dynamics in the case of a small polydisperse system at low densities. The discrete-time dynamics we consider is made up only by cSwap moves, the positions of the disks $\br^N$ being fixed at all times. The state of the system $\mathcal{C}=\{\sigma^N,i\}$ is specified by assigning a diameter to each of the $N$ particles --the resulting permutation of diameters is denoted by $\sigma^N$-- and by the lifting degree of freedom $i$, i.e. the label of the active particle. The system can access a subset of $N!\times N$ configurations. The configurations in the subset satisfy the non-overlapping hard disks condition. The probability $\pi_t(\mC)$ for the system to be in state $\mC$ at time $t$ obeys the discrete time Markov dynamics
\begin{equation}
    \pi_{t+1}(\mC) = \sum_{\mC'} P(\mC'\to\mC) \pi_t(\mC') ,
\end{equation}
with $P$ the transition matrix encoding the probability to jump from one configuration to another during a discrete time-step. The convergence properties of the dynamics are encoded in the spectrum of $P$. Proving that the cSwap dynamics is ergodic amounts to showing that the spectrum of $P$ has a unique, nondegenerate, eigenvalue $\lambda=1$ lying on the unit circle, and that all the other eigenvalues have a norm strictly smaller than $1$~\cite{newman1999monte}. 

We consider the case where $N=4$ and the particles are far away from each other, so that no overlap between them can be generated upon permutation of the radii. To write down the transition matrix describing the cSwap dynamics, we decompose the configuration space into a tensor product $S_4 \otimes i$, where $S_4$ is an element of the permutation group of 4 elements, and $i$ is the lifting degree of freedom labeling the active, expanding particle. The permutations are labeled by natural numbers in the following way:
\begin{equation}
\begin{array}{cccc}
    1 \equiv \{1,2,3,4\} & 2 \equiv \{2,3,4,1\} & 3 \equiv \{3,4,1,2\} & 4 \equiv \{4,1,2,3\}\\
    5 \equiv \{1,3,4,2\} & 6 \equiv \{3,4,2,1\} & 7 \equiv \{4,2,1,3\} & 8 \equiv \{2,1,3,4\}\\
    9 \equiv \{1,2,4,3\} & 10 \equiv \{2,4,3,1\} & 11 \equiv \{4,3,1,2\} & 12 \equiv \{3,1,2,4\}\\
    13 \equiv \{2,4,1,3\} & 14 \equiv \{4,1,3,2\} & 15 \equiv \{1,3,2,4\} & 16 \equiv \{3,2,4,1\}\\
    17 \equiv \{2,3,1,4\} & 18 \equiv \{3,1,4,2\} & 19 \equiv \{1,4,2,3\} & 20 \equiv \{4,2,3,1\}\\
    21 \equiv \{2,1,4,3\} & 22 \equiv \{1,4,3,2\} & 23 \equiv \{4,3,2,1\} & 24 \equiv \{3,2,1,4\}\\
\end{array}
\end{equation}
Here the numbers inside the brackets denote the index of the particles, ordered according to their diameters, from the smallest to the largest. For example, $\{2,3,4,1\}$ is the configurations where particle with label $1$ has the largest diameter, particle with label $4$ has the second largest diameter, and so on. If the system is in a configuration where particle $3$ is the active one, we would have $\mC = \{2,3,4,1\}\otimes 3$.

With this notation, the transition matrix can be written as a $24\times24$ block matrix, each block being made by a $4\times4$ sub-block. Transitions between blocks represent changes in the diameter assignments for the different particles, while transitions within a block represent change in the active degree of freedom. The transition matrix describing the cSwap dynamics is
\begin{equation}\label{eq:P4}
P= \left[\begin{array}{*{24}c}
D_4 & B_1 & \cdot & \cdot & B_2 & \cdot & \cdot & \cdot & B_3 & \cdot & \cdot & \cdot & \cdot & \cdot & \cdot & \cdot & \cdot & \cdot & \cdot & \cdot & \cdot & \cdot & \cdot & \cdot \\
\cdot & D_4 & B_1 & \cdot & \cdot & \cdot & \cdot & \cdot & \cdot & \cdot & \cdot & \cdot & B_2 & \cdot & \cdot & \cdot & B_3 & \cdot & \cdot & \cdot & \cdot & \cdot & \cdot & \cdot \\
\cdot & \cdot & D_4 & B_1 & \cdot & B_3 & \cdot & \cdot & \cdot & \cdot & \cdot & B_2 & \cdot & \cdot & \cdot & \cdot & \cdot & \cdot & \cdot & \cdot & \cdot & \cdot & \cdot & \cdot \\
B_1 & \cdot & \cdot & D_4 & \cdot & \cdot & \cdot & \cdot & \cdot & \cdot & \cdot & \cdot & \cdot & B_3 & \cdot & \cdot & \cdot & \cdot & \cdot & B_2 & \cdot & \cdot & \cdot & \cdot \\
\cdot & \cdot & \cdot & \cdot & D_4 & B_1 & \cdot & \cdot & \cdot & \cdot & \cdot & \cdot & \cdot & \cdot & B_3 & \cdot & \cdot & \cdot & B_2 & \cdot & \cdot & \cdot & \cdot & \cdot \\
\cdot & \cdot & B_3 & \cdot & \cdot & D_4 & B_1 & \cdot & \cdot & \cdot & \cdot & \cdot & \cdot & \cdot & \cdot & \cdot & \cdot & \cdot & \cdot & \cdot & \cdot & \cdot & \cdot & B_2 \\
\cdot & \cdot & \cdot & \cdot & \cdot & \cdot & D_4 & B_1 & \cdot & \cdot & \cdot & \cdot & \cdot & B_2 & \cdot & \cdot & \cdot & \cdot & \cdot & B_3 & \cdot & \cdot & \cdot & \cdot \\
\cdot & B_2 & \cdot & \cdot & B_1 & \cdot & \cdot & D_4 & \cdot & \cdot & \cdot & \cdot & \cdot & \cdot & \cdot & \cdot & \cdot & \cdot & \cdot & \cdot & B_3 & \cdot & \cdot & \cdot \\
B_3 & \cdot & \cdot & \cdot & \cdot & \cdot & \cdot & \cdot & D_4 & B_1 & \cdot & \cdot & \cdot & \cdot & \cdot & \cdot & \cdot & \cdot & \cdot & \cdot & \cdot & B_2 & \cdot & \cdot \\
\cdot & \cdot & \cdot & \cdot & \cdot & \cdot & \cdot & \cdot & \cdot & D_4 & B_1 & \cdot & B_3 & \cdot & \cdot & \cdot & B_2 & \cdot & \cdot & \cdot & \cdot & \cdot & \cdot & \cdot \\
\cdot & \cdot & \cdot & B_2 & \cdot & \cdot & \cdot & \cdot & \cdot & \cdot & D_4 & B_1 & \cdot & \cdot & \cdot & \cdot & \cdot & \cdot & \cdot & \cdot & \cdot & \cdot & B_3 & \cdot \\
\cdot & \cdot & \cdot & \cdot & \cdot & \cdot & \cdot & \cdot & B_1 & \cdot & \cdot & D_4 & \cdot & \cdot & \cdot & B_2 & \cdot & B_3 & \cdot & \cdot & \cdot & \cdot & \cdot & \cdot \\
\cdot & \cdot & \cdot & \cdot & \cdot & \cdot & \cdot & B_2 & \cdot & B_3 & \cdot & \cdot & D_4 & B_1 & \cdot & \cdot & \cdot & \cdot & \cdot & \cdot & \cdot & \cdot & \cdot & \cdot \\
\cdot & \cdot & \cdot & B_3 & \cdot & \cdot & \cdot & \cdot & \cdot & \cdot & \cdot & \cdot & \cdot & D_4 & B_1 & \cdot & \cdot & \cdot & \cdot & \cdot & \cdot & \cdot & B_2 & \cdot \\
\cdot & \cdot & \cdot & \cdot & B_3 & \cdot & \cdot & \cdot & B_2 & \cdot & \cdot & \cdot & \cdot & \cdot & D_4 & B_1 & \cdot & \cdot & \cdot & \cdot & \cdot & \cdot & \cdot & \cdot \\
\cdot & \cdot & B_2 & \cdot & \cdot & \cdot & \cdot & \cdot & \cdot & \cdot & \cdot & \cdot & B_1 & \cdot & \cdot & D_4 & \cdot & \cdot & \cdot & \cdot & \cdot & \cdot & \cdot & B_3 \\
\cdot & B_3 & \cdot & \cdot & \cdot & \cdot & \cdot & \cdot & \cdot & \cdot & \cdot & \cdot & \cdot & \cdot & \cdot & \cdot & D_4 & B_1 & \cdot & \cdot & B_2 & \cdot & \cdot & \cdot \\
\cdot & \cdot & \cdot & \cdot & \cdot & B_2 & \cdot & \cdot & \cdot & \cdot & \cdot & B_3 & \cdot & \cdot & \cdot & \cdot & \cdot & D_4 & B_1 & \cdot & \cdot & \cdot & \cdot & \cdot \\
B_2 & \cdot & \cdot & \cdot & \cdot & \cdot & \cdot & \cdot & \cdot & \cdot & \cdot & \cdot & \cdot & \cdot & \cdot & \cdot & \cdot & \cdot & D_4 & B_1 & \cdot & B_3 & \cdot & \cdot \\
\cdot & \cdot & \cdot & \cdot & \cdot & \cdot & B_3 & \cdot & \cdot & \cdot & B_2 & \cdot & \cdot & \cdot & \cdot & \cdot & B_1 & \cdot & \cdot & D_4 & \cdot & \cdot & \cdot & \cdot \\
\cdot & \cdot & \cdot & \cdot & \cdot & \cdot & \cdot & B_3 & \cdot & B_2 & \cdot & \cdot & \cdot & \cdot & \cdot & \cdot & \cdot & \cdot & \cdot & \cdot & D_4 & B_1 & \cdot & \cdot \\
\cdot & \cdot & \cdot & \cdot & \cdot & \cdot & \cdot & \cdot & \cdot & \cdot & \cdot & \cdot & \cdot & \cdot & B_2 & \cdot & \cdot & \cdot & B_3 & \cdot & \cdot & D_4 & B_1 & \cdot \\
\cdot & \cdot & \cdot & \cdot & \cdot & \cdot & B_2 & \cdot & \cdot & \cdot & B_3 & \cdot & \cdot & \cdot & \cdot & \cdot & \cdot & \cdot & \cdot & \cdot & \cdot & \cdot & D_4 & B_1 \\
\cdot & \cdot & \cdot & \cdot & \cdot & \cdot & \cdot & \cdot & \cdot & \cdot & \cdot & \cdot & \cdot & \cdot & \cdot & B_3 & \cdot & B_2 & \cdot & \cdot & B_1 & \cdot & \cdot & D_4 \\
\end{array}\right].
\end{equation}
The dot symbol $\cdot$ denote $4\times 4$ matrices whose entries are all zeros. The diagonal block is a sum of two terms, $D_4 \equiv A_4+B_4$. The block of type $A_4$ encodes transitions involving only the active degree of freedom, 
\begin{equation}
    A_4 \equiv \begin{bmatrix} 
    \alpha & \alpha & \alpha & \alpha \\
    \alpha & \alpha & \alpha & \alpha \\
    \alpha & \alpha & \alpha & \alpha \\
    \alpha & \alpha & \alpha & \alpha\\ 
    \end{bmatrix},
\end{equation}
with $\alpha=1/4^2 = 1/16$. They are associated to random resampling of the activity label, and they are used here to ensure that $P$ is ergodic. The blocks $B_i$ encode transitions generated by the cSwap moves, involving an inflation of the active particle and the cascade of swaps. They are given by: 
\begin{equation}
    \begin{array}{cccc}
    B_1 \equiv 
        \begin{bmatrix} 
        \circ &\circ &\circ & 1-\frac{1}{4} \\ 
        \circ &\circ &\circ & \circ \\ 
        \circ &\circ & \circ &\circ \\ 
        \circ &\circ & \circ &\circ \end{bmatrix} & 
    B_2 \equiv 
        \begin{bmatrix} 
            \circ &\circ &\circ & \circ \\ 
            1-\frac{1}{4} &\circ &\circ & \circ \\ 
            \circ &\circ & \circ &\circ \\ 
            \circ &\circ & \circ &\circ 
        \end{bmatrix} & 
    B_3 \equiv 
        \begin{bmatrix} 
            \circ &\circ &\circ & \circ \\ 
            \circ &\circ &\circ & \circ \\ 
            \circ &1-\frac{1}{4} & \circ &\circ \\ 
            \circ &\circ & \circ &\circ 
        \end{bmatrix} &
    B_4 \equiv 
        \begin{bmatrix} 
            \circ &\circ &\circ & \circ \\ 
            \circ &\circ &\circ & \circ \\ 
            \circ &\circ & \circ &\circ \\ 
            \circ &\circ & 1-\frac{1}{4} &\circ 
        \end{bmatrix},
    \end{array}
\end{equation}
with the symbol `$\circ$' denoting the entries with value $0$.

\begin{figure}[h]
    \includegraphics[width=0.5\textwidth]{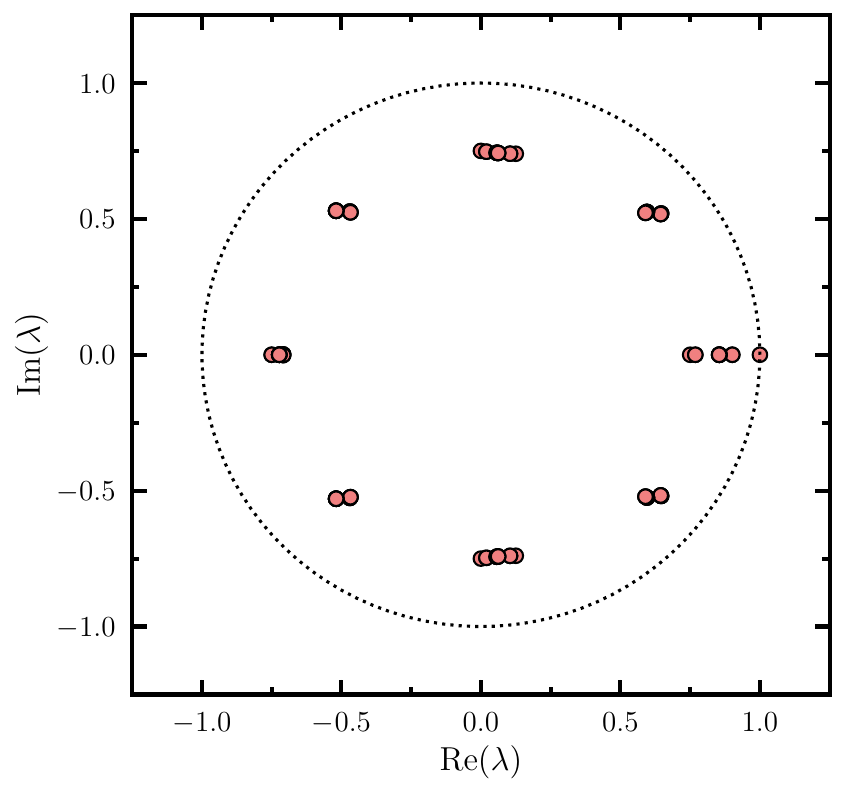}
    \caption{Representation in the complex plane of the eigenvalues $\lambda$ of the transition matrix $P$ given by Eq.~\eqref{eq:P4} for the cSwap dynamics in the case of a system of $N=4$ particles. The dotted line is the unit circle.}\label{fig:spectrum}
\end{figure}

The matrix $P$ is doubly stochastic, i.e. the sum of all the elements belonging to a fixed row, or fixed column, is one. A stationary solution of the stochastic process associated with $P$ is given by
\begin{equation}
    \pi_{ss} \equiv \frac{1}{4\times 4!} \bigoplus_{i=1}^{24} \left[1,1,1,1 \right]^T, 
\end{equation}    
which is the tensor product of the Boltzmann distribution for hard disks times a uniform distribution of the active degree of freedom among the $4$ particles. To show ergodicity, we inspect the spectrum of $P$. Its eigenvalues $\lambda$ are shown in the complex plane in Fig.~\ref{fig:spectrum}.
The only eigenvalue lying on the unit circle is $\lambda=1$, thus proving the ergodicity of the cSwap Markov chain. We observe that the other eigenvalues tend to accumulate at the vertices of an octagon inside the unit circle. This is a consequence of the introduction of a refreshment probability for the label of the active particle. Without such a refreshment dynamics, the Markov chain would be periodic, and there would be $2N$ eigenvalues lying on the unit circle. The introduction of a refreshment rate pushes the eigenvalues inside the unit circle, making the Markov Chain aperiodic. 

\section{Additional numerical evidence of ergodicity for cSwap dynamics}

\begin{figure}[t]
    \includegraphics[width=\textwidth]{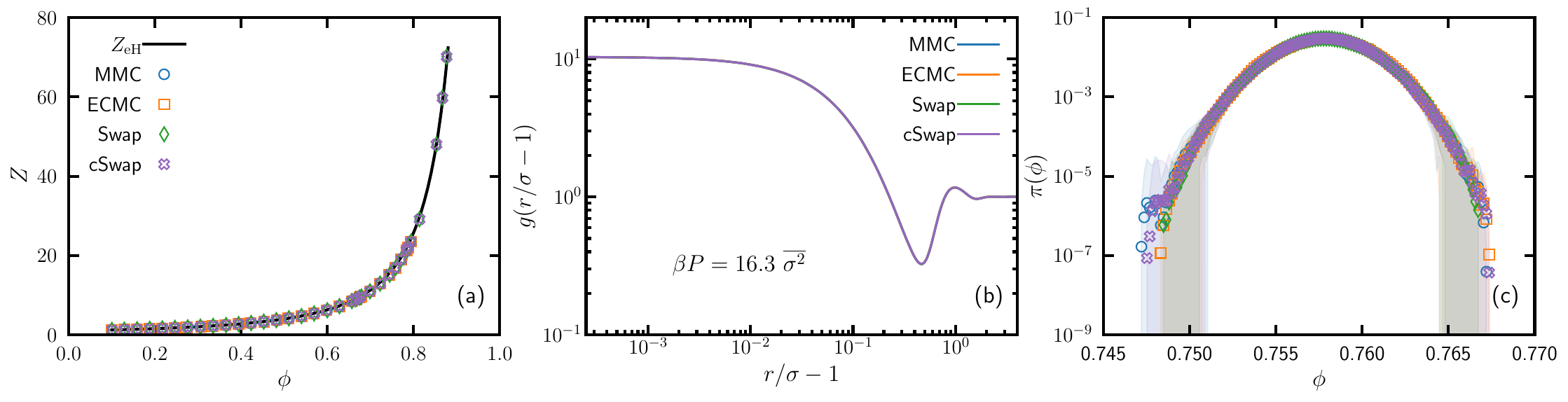}
    \caption{Numerical test of ergodicity for the cSwap and ECMC algorithms. (a) Equation of state $Z(\phi)$ for the polydisperse hard disks system. (b) Rescaled radial distribution function $g(r/\sigma - 1)$ as a function of the distance from its first peak $r/\sigma - 1$. (c) Probability distribution function of the packing fractions $\pi(\phi)$ explored by the system during $NPT$ simulations at $\beta P=16.3 \overline{\sigma^2}$.} 
    \label{fig:ergodicitytest}
\end{figure}

We present numerical tests supporting the ergodicity of the cSwap and ECMC algorithms. The first test is the numerical calculation of the equation of state $Z(\phi)$ of the polydisperse system considered in the main text. The equation of state relates the reduced pressure $Z=\frac{\beta P}{\rho}$ to the packing fraction $\phi$. Here $P$ is the pressure applied to the system, $\rho$ is the number density and $\beta^{-1}=k_\text{\tiny{B}}T$. It can be obtained from simulations in the $NPT$ ensemble, where one has access to the running averages of $Z$ and $\phi$ at fixed applied pressure $P$. The results are shown in Fig.~\ref{fig:ergodicitytest}(a),
where they are compared with an extension to polydisperse systems of the empirical Henderson formula \cite{henderson1975simple, santos1999equation} which reads 
\begin{equation}\label{eq:santos}
    \begin{split}
        Z_{\rm eH} &= \frac{1 - \left(1 - \frac{\overline{\sigma}^2}{\overline{\sigma^2}}\right)\phi + (b-3)\frac{\overline{\sigma}^2}{\overline{\sigma^2}}\phi^2}{\left(1 - \phi\right)^2} , \\
        b &\equiv \frac{16}{3} - \frac{4\sqrt{3}}{\pi}.
    \end{split}
\end{equation}
The agreement with the empirical formula and with the conventional Metropolis and Swap algorithms is excellent, and this serves as a guide to detect deviations from one algorithm to another. We find that all algorithms agree with each other. 

We next compare the rescaled radial distribution function $g(r/\sigma)$ in the NPT ensemble in Fig.~\ref{fig:ergodicitytest}(b). Its expression is given by
\begin{equation}\label{eq:gr}
    \begin{split}    
        g(x) &\equiv \sum_{\substack{i,j\\ i<j}} C_{ij} \int_{x_b}^{x_b+\Delta x} \delta\left( x' - \frac{r_{ij}}{\sigma_{ij}} \right)dx' \\
        C_{ij} &\equiv \frac{1}{2\pi (x_b+\Delta x/2)\Delta x \sigma_{ij}^2\rho(N-1)}
    \end{split}
\end{equation}
where $\sigma_{ij} \equiv \frac{1}{2}\left(\sigma_i+\sigma_j\right)$, $\rho=N/L^2$ is the number density of the system, and we collect the rescaled interparticle distances in a histogram with bin width $\Delta x$. $x_b= b\Delta_x$ is the coordinate of the $b$-th bin, with $b$ chosen so that $x_b \leq x< x_b +\Delta x$. Again, the curves obtained with the different algorithms superimpose on each other. 

Finally, we report in Fig.~\ref{fig:ergodicitytest}(c) the histogram of the packing fractions explored during an NPT simulation, $\pi(\phi)$, for a fixed pressure $P$. Again all algorithms explore the same fluctuations, including in the tails of the distribution, showing that the same Boltzmann distribution is indeed properly sampled in all our algorithms. 

\section{Choice of time units for the different Monte Carlo algorithms}

\begin{figure}[h]
    \includegraphics[width=0.6\textwidth]{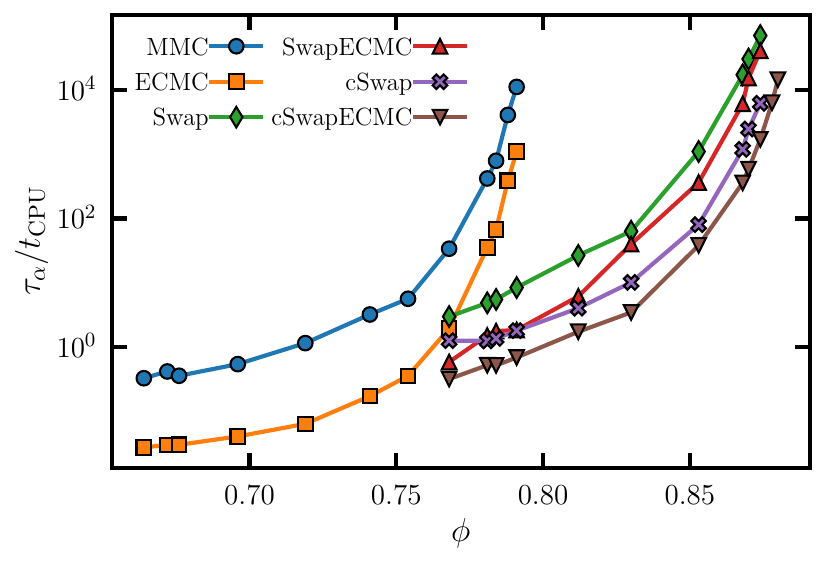}
    \caption{Equilibrium relaxation times of the six algorithms investigated in the main text, in units of CPU time. Here $t_\text{CPU}=1$ second.}
    \label{fig:taualphacpu}
\end{figure}

In the main text a unit of time $t_\text{move}$ was chosen based upon the elementary transitions involved in each algorithm. Here, we clarify its relation with the CPU time required by the algorithms studied in the main text. We first observe that in all the algorithms investigated, during a unit of $t_\text{move}$ only one evaluation of the overlap of a given disk with its neighbors is performed. This is usually the most computationally demanding task in hard disks simulations. We thus expect the trend of the relaxation time in units of CPU time to resemble the one obtained using units of $t_\text{move}$. This is confirmed in Fig.~\ref{fig:taualphacpu}, where we show the relaxation time $\tau_\alpha$, now measured in units of CPU time, as a function of $\phi$. The trend of the curves is virtually identical to the one shown in Fig.~1 of the main text. 

\begin{figure}[h]
    \includegraphics[width=\textwidth]{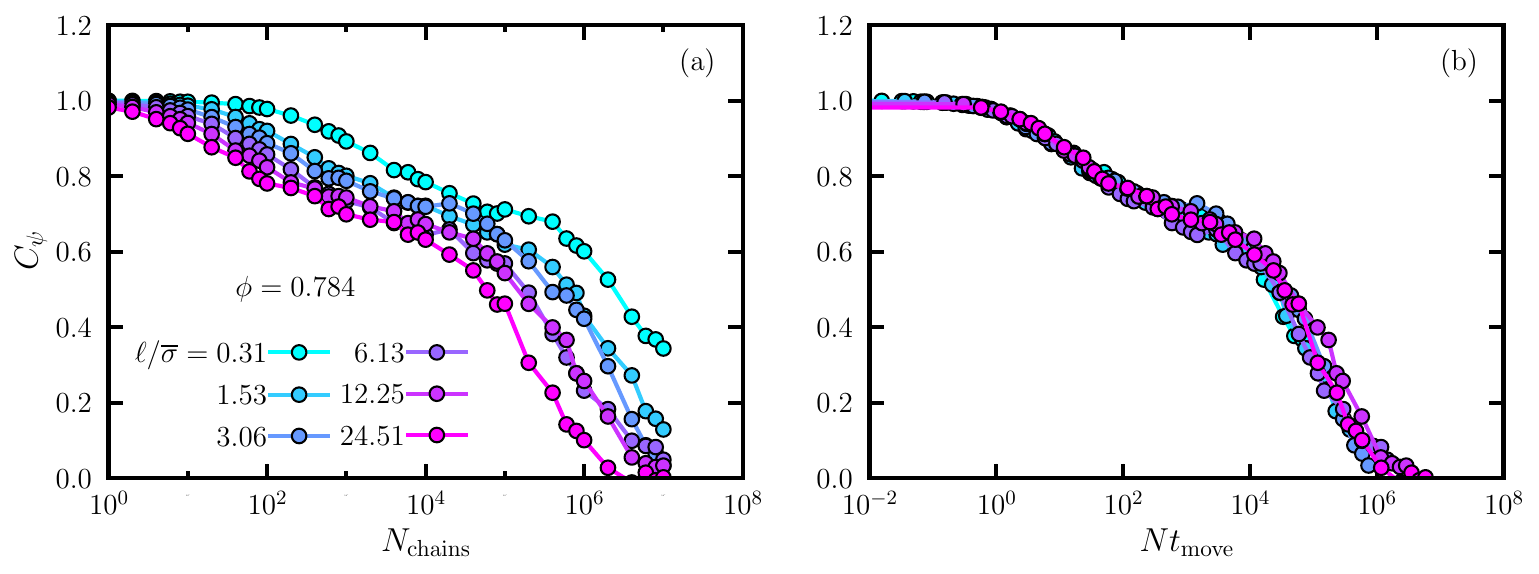}
    \caption{Equilibrium correlation function for the ECMC dynamics, using different values of the chain length $\ell$. Time is measured in units of (a) the number of chains displaced, and (b) $N t_\text{move}$.}
    \label{fig:ecmc_nchains}
\end{figure}

When using the ECMC algorithm, it is also possible to measure times using the number of directed chains of particles that have moved~\cite{bernard2009event}. When comparing ECMC with other types of algorithms, however, this choice is a poor indicator of its efficiency, as it hides the number of particles--and hence of event determinations--that are involved in each chain. This is demonstrated in Fig.~\ref{fig:ecmc_nchains}(a), where the time relaxation of $C_\psi$ is shown as a function the number of chain displaced during an ECMC in the $NVT$ ensemble, for different values of the chain length $\ell$. When times are measured according to the number of chains displaced, longer chains have a stronger impact on the system relaxation. However, if one measures times in units of $N t_\text{move}$, thus counting individual particle displacements, we see that all the relaxation curves now collapse. This collapse implies that the efficiency of ECMC (in CPU time) is nearly independent of $\ell$.

\section{Efficiency of the algorithms as a function of system size}

In this section we address the question of how the gain provided by SwapECMC, cSwap and cSwapECMC behaves with the size of the system. We study the relaxation dynamics for polydisperse systems of $N=1024$, 2048, and 4096 hard disks, with the polydispersity defined as in Sec.~\ref{sec:different_model}. To avoid uncontrolled fluctuations in the distribution of the diameters and make a clear comparison between different system sizes, we generate the diameters $\{\sigma_i\}_{i=1,\ldots,N}$ in each system of $N$ particles in the following way: we first take $N$ numbers $a_i = i/N$ with a uniform spacing in the interval $[0,1]$. From each number $a_i$, the diameter $\sigma_i$ is generated using the following relation
\begin{equation}
    \sigma_i = \frac{\sigma_\text{max}}{\sqrt{1 - a_i + a_i \left(\frac{\sigma_\text{max}}{\sigma_\text{min}}\right)^2}}, 
\end{equation}
which maps a random number generated from the uniform distribution in the interval $[0,1]$ to a random number generated from a power law distribution $\propto \sigma^{-3}$ between $\sigma_\text{min}$ and $\sigma_\text{max}$. 

\begin{figure}
    \includegraphics[width=0.6\textwidth]{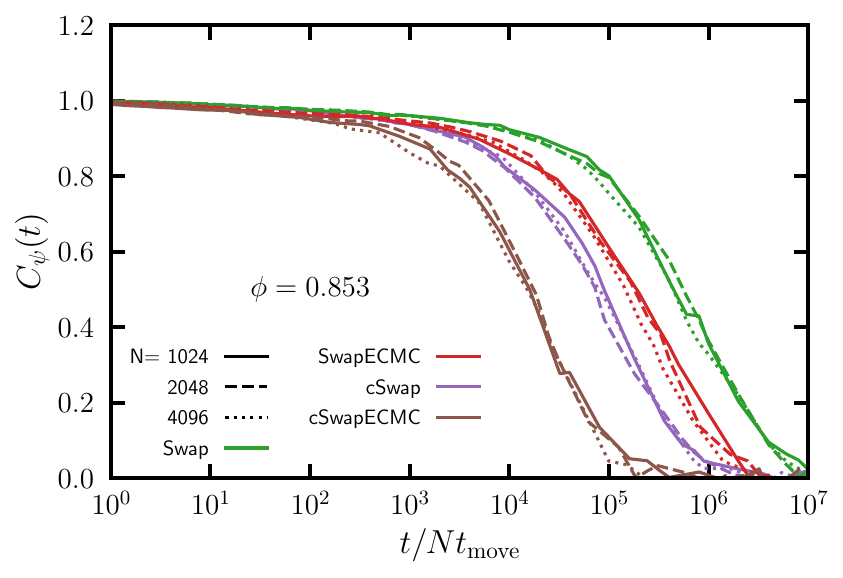}
    \caption{\label{fig:size_effects} Time dependence of the hexatic correlation function for systems of different sizes using the Swap, SwapECMC, cSwap and cSwapECMC algorithms in the NVT ensemble.}
\end{figure}

We run NVT simulations using the Swap, SwapECMC, cSwap, cSwapECMC algorithms for the three system sizes at a high packing fraction $\phi=0.853$, and we track the decay of the correlation function $C_\psi(t)$. The resulting curves, displayed in Fig~\ref{fig:size_effects}, demonstrate that the gain provided by the different algorithms is constant with respect to the size of the system. This is in line with previous results regarding swap efficiency in glass-formers. 

\bibliography{biblio-cswap}